\documentclass[prc,preprint,showpacs,showkeys,nofootinbib,tightenlines]{revtex4}
\usepackage{graphicx}  %
\usepackage{bm}  %                                          

\newcommand{\simge}{\hspace*{0.2em}\raisebox{0.5ex}{$>$}
     \hspace{-0.8em}\raisebox{-0.3em}{$\sim$}\hspace*{0.2em}}
\newcommand{\simle}{\hspace*{0.2em}\raisebox{0.5ex}{$<$}
     \hspace{-0.8em}\raisebox{-0.3em}{$\sim$}\hspace*{0.2em}}
\def\dpst{\displaystyle}

\newcommand{\order}[1]{{\cal O}(#1)}

\begin{document}

\title{The Hypertriton in Effective Field Theory}

\author{H.-W. Hammer}\email{hammer@mps.ohio-state.edu}

\affiliation{Department of Physics,
         The Ohio State University, Columbus, OH\ 43210, USA}

%\date{\today}
\date{October 15, 2001}

\begin{abstract}
Doublet $\Lambda d$ scattering and the hypertriton are studied in 
the framework of an effective field theory for large scattering 
lengths.  As in the triton case,
consistent renormalization requires a one-parameter
three-body force at leading order whose renormalization group evolution
is governed by a limit cycle. Constraining unknown parameters from 
symmetry considerations and the measured binding energy of the hypertriton,
we calculate the low-energy phase shifts for doublet $\Lambda d$ scattering.
For the low-energy parameters, we find $a_{\Lambda d}=(16.8^{+4.4}_{-2.4})$ 
fm and $r_{\Lambda d}=(2.3 \pm 0.3)$ fm, where the errors are due to the
uncertainty in the hypertriton binding energy. 
Since the hypertriton is extremely shallow, low-energy three-body 
observables in this channel are very insensitive to the exact values of 
the $\Lambda N$ low-energy parameters.
\end{abstract}

\smallskip
\maketitle  

\section{Introduction}
\label{sec:intro}

In recent years, there has been much interest in applying
Effective Field Theory (EFT) methods to nuclear systems with
two or more nucleons \cite{Weinberg,NN98,NN99,Birareview,Bea00}.
EFT's provide a powerful framework to explore a separation of scales 
in physical systems in order to perform systematic, model-independent
calculations \cite{EFT}.  If, for example, the momenta $k$ of two
particles are much smaller than the inverse range of 
their interaction $1/R$, observables can be expanded in powers of $kR$.
All short-distance effects are systematically absorbed into a few
low-energy constants using renormalization. The EFT approach allows
for accurate calculation of low-energy processes with well-defined
error estimates. For low-energy nuclear few-body systems, the 
long-distance scale is set by the large two-body scattering lengths, 
while the short distance scale is set by the range of the nuclear force
or the inverse pion mass. In an EFT, the long-distance physics is included
explicitly, while the corrections from short-distance physics
are calculated in an expansion in the ratio of these two scales.

For very low energies and momenta ($p \simle m_\pi$), even pion
exchange can be considered \lq\lq short-distance'' physics. In this case,
one can use an effective Lagrangian including only contact interactions.
The large S-wave scattering lengths require that the leading two-body 
contact interaction is treated nonperturbatively \cite{vKo99,KSW98}.
In the two-body system, this program has been very successful 
(see e.g. Refs.~\cite{Bea00,CRS99,BeS00} and references therein). 

In the nuclear three-body system, considerable progress 
has been made as well 
\cite{BeK98,BHK98,BHK99,BeG00,GBG00,RuK01,BHK00,BlG00,HaM01,HaM01b}.
Most of the work has concentrated on the neutron-deuteron system. 
In channels where the Pauli principle or centrifugal barrier suppress
sensitivity to short-distance physics, the EFT can be extended in a 
straightforward way \cite{BeK98,BHK98,BeG00,GBG00}. Recently,
Coulomb repulsion has been included and the phase shifts for
proton-deuteron scattering in the spin quartet ($J=3/2$) have been 
calculated \cite{RuK01}.
Most interesting, however, is the S-wave in doublet ($J=1/2$)
neutron-deuteron scattering which displays a number of surprising
phenomena such as the Thomas and Efimov effects or the Phillips
line \cite{Tho35,Phi68,Efi71}.  In this channel, the
renormalization requires a one-parameter three-body force at 
leading order whose renormalization group evolution is governed by a 
limit cycle  \cite{BHK99,BHK00}. Variation of this three-body force
gives a compelling explanation of the Phillips line \cite{Phi68}.
Recently, the effective range corrections to S-wave neutron-deuteron 
scattering in the doublet channel have been calculated \cite{HaM01b}. 
At this order, good agreement with available phase shift analyses and
potential model calculations is obtained.

The purpose of the present paper is to study doublet $\Lambda d$ scattering
and the hypertriton using EFT methods. The hypertriton is the 
lightest hypernucleus and consists of a triton with one neutron
replaced by a $\Lambda$. It is the simplest strange three-body halo
nucleus with a separation energy into a deuteron and a $\Lambda$ 
of only $B^\Lambda = 0.13 \pm 0.05$ MeV \cite{Jur73,Dav91}. 
The total binding energy is $B_3^\Lambda=2.35$ MeV. The hypertriton has 
been  studied extensively using various hyperon-nucleon potentials
(see e.g. Refs.~\cite{ClL85,AfG89,Con92,MiG93,CJF97,FeJ01} and
references therein). The most sophisticated calculations
include both tensor forces and $\Lambda \leftrightarrow \Sigma$ conversion 
effects \cite{AfG89}. Obtaining the correct binding energy for the 
hypertriton in these potential models is delicate, which has
motivated the use of this system to study details of the hyperon-nucleon 
interaction \cite{CJF97}.

The EFT framework offers a different perspective and stresses
the universal aspects of the problem. 
In this paper, we study $\Lambda d$ scattering and the hypertriton
in the framework of an EFT for three-body systems with large two-body 
scattering lengths \cite{BHK99,BHK00}. The small $N-\Lambda$ mass 
difference can be exploited by expanding in the parameter
$y=(M_\Lambda-M)/(M_\Lambda+M)\approx 0.1$. After this expansion is
performed, the hypertriton emerges naturally as a shallow Efimov state. 
To leading order, both $NN$ and $\Lambda N$ S-waves will contribute
to the hypertriton. While the $NN$ S-waves have a large scattering
length, the $\Lambda N$ scattering lengths are believed to be natural.
Therefore, it is not obvious that an EFT for large two-body scattering 
lengths is applicable. In the following, we motivate the use of this
EFT.

Unfortunately, there are only few $\Lambda N$ scattering data and these
are at relatively high energies. Model independent analyses using
effective range theory indicate mainly S-wave scattering, but 
afford many solutions for the low-energy parameters 
and are essentially inconclusive \cite{Ale68,Sec68}. 
For example in Ref.~\cite{Sec68}, the extracted one-standard-deviation
bounds were $0> a_1 > -15$ fm, $0 < r_1 < 15$ fm, $-0.6$ fm  $> a_3 > -3.2$ fm,
and $2.5$ fm $< r_3 < 15$ fm where $a_1$ ($a_3$) and $r_1$ ($r_3$) 
are the singlet (triplet) scattering length and effective range,
respectively. These bounds do not exclude the case of large 
scattering lengths.

Various sets of low-energy parameters are also known from extrapolations 
of the few higher energy data using hyperon-nucleon potentials 
\cite{NRS73,RSY99,CJF97,HHS89,RHS94}. The extracted 
$\Lambda N$ scattering lengths and effective ranges are
generally of natural size [$\order{1/m_\pi}$]. Both the 
$^3S_1$ and $^1S_0$ $\Lambda N$ partial waves have a virtual bound
state. The pole momenta extracted from the low-energy extrapolations
are $\sim 60$ MeV for the $^1S_0$ and $\sim 70$ MeV for the $^3S_1$
partial wave. However, it is not clear how large the errors and
model-dependence in these low-energy extrapolations are. 

An EFT analysis of hyperon nucleon scattering and hyperon mass
shifts in the nuclear medium was recently performed in Ref.~\cite{KDT01}.
The $\Lambda N$ scattering lengths, however, were used as input in this 
analysis.

Because the hypertriton is so weakly bound, it should only be sensitive 
to the extreme long-distance properties of the $\Lambda N$ interaction. 
The typical momentum scale of the hypertriton can be estimated from
its binding energy via
\begin{equation}
\label{eq:3bdymom}
\gamma_3^\Lambda \sim 2\sqrt{(MB_3^\Lambda-\gamma_d^2)/3}\approx 0.3 
\gamma_d\,,
\end{equation}
where $M=938.9$ MeV is the nucleon mass and $\gamma_d=45.68$ MeV the 
deuteron pole momentum. Since $\gamma_3^\Lambda$ is so small, the
contribution from the effective ranges will be suppressed even if 
the $\Lambda N$ low-energy parameters are $\order{1/m_\pi}$
as extracted from the potential models (see Section \ref{sec:res} 
for a more quantitative estimate). Consequently, the 
use of the EFT for large two-body scattering lengths is justified.

In the next Section, we write down the effective Lagrangian and
discuss the power counting of the EFT. We then review
the EFT for the $NN$ and $\Lambda N$ two-body subsystems. In Section
\ref{sec:3bdy}, we obtain the three-body integral equations for the
$\Lambda NN$ system and in Section \ref{sec:res} we present our
results and conclusions. Some technical details are given in the
Appendices.

\section{Two-Body System}
\label{sec:2body}

The power counting in an EFT is determined by the physical 
scales in the system under consideration. For the
hypertriton, we are only interested in very low-energies, 
where all physics (even pion exchange) can be considered 
\lq\lq short-range''. The two-body subsystems ($NN$ and $\Lambda N$) 
are characterized by two scales: a long-distance scale (the large S-wave 
scattering length) and a short-distance scale that is given by the 
longest-range physics excluded from the EFT. For an EFT without pions, 
one expects the short-distance scale $R$ to be of $\order{1/m_\pi}$.
However, since the $\Lambda\;(I=0)$ and nucleon ($I=1/2$) cannot exchange 
a pion with $I=1$ and conserve isospin, the short-distance scale for the 
hypertriton should be set by the two-pion exchange \cite{AfG89}. 
A power counting to deal with unnaturally large two-body scattering 
lengths has been proposed in Refs.~\cite{vKo99,KSW98}. This counting
takes $p\sim 1/a \sim Q$, where $p$ is the typical momentum
and $1/a$ is the inverse scattering length. The pole or binding momentum
$\gamma$ is of $\order{Q}$ as well since $\gamma=1/a +\order{r/a^2}$
with $r\sim R$ the effective range. The expansion of the
EFT is in powers of $Q R \approx \gamma r$ (see Refs.~\cite{GBG00,HaM01b}
for more details). 

For the study of three-body systems, it is convenient to employ
the dibaryon formalism \cite{Kap97}, where an auxiliary field is
used to describe two baryons interacting in a given partial wave. 
We write down an effective Lagrangian including nucleons, $\Lambda$'s,
and dibaryon fields for the deuteron ($d$) as well as the $^3S_1$ 
($u^3$) and $^1S_0$ ($u^1$) $\Lambda N$ partial waves,
\begin{eqnarray}
\label{lagd}
{\cal L}&=&N^\dagger \left(i\partial_t +\frac{\vec{\nabla}^2}{2M}\right)N
+\Lambda^\dagger \left(i\partial_t +\frac{\vec{\nabla}^2}{2M_\Lambda}\right)
\Lambda\\
& &+\Delta_d d_l^\dagger d_l-\frac{g_d}{2}\left[ d_l^\dagger N^T 
(i\tau_2)(i\sigma_l \sigma_2) N + \mbox{H.c.} \right] \nonumber \\
& &+\Delta_{3} (u^3_l)^\dagger u^3_l-ig_3 \left[ (u^3_l)^\dagger \Lambda^T 
(i\sigma_l \sigma_2) N + \mbox{H.c.} \right] \nonumber\\
& &+\Delta_{1} (u^1)^\dagger u^1-ig_1 \left[ (u^1)^\dagger \Lambda^T 
(i\sigma_2) N + \mbox{H.c.} \right]+\ldots\,, \nonumber
\end{eqnarray}
where H.c. denotes the Hermitian conjugate and
the dots indicate terms with more derivatives and/or fields.
The terms with more derivatives are suppressed at low energies,
while four- and higher-body forces do not contribute. Three-body
terms will be addressed in the next Section. In Eq.~(\ref{lagd}),
indices occuring repeatedly are summed over and 
$\sigma_i$ ($\tau_i$) are the usual Pauli matrices acting in spin
(isospin) space. Spinor and isospinor indices are suppressed.
The parameters $\Delta$ and $g$ are not independent and only the 
combination $\Delta/g^2$ enters physical observables. 

The Lagrangian 
(\ref{lagd}) describes low-energy S-wave scattering of nucleons and 
$\Lambda$'s and reproduces the effective range expansion 
up to the scattering length term, which is sufficient for
our purposes. To go to higher orders in $QR$, the two-body effective 
range can be included  with a kinetic term for the auxiliary field 
\cite{Kap97}.
Only the partial waves contributing in the hypertriton channel to leading 
order in $QR$ are included in the effective Lagrangian (\ref{lagd}).
The $^1S_0$ $NN$ amplitude does not contribute because of isospin,
while higher partial waves and the tensor force are suppressed by at 
least two powers of the expansion parameter $Q R$ \cite{GBG00}. 
After integrating out the auxiliary fields $d$, $u^3$, and $u^1$,
it is straightforward to show that the Lagrangian (\ref{lagd})
is equivalent to a Lagrangian with nucleon and $\Lambda$
fields only (cf. Refs.~\cite{BHK99,BHK00}).

\begin{figure}[t]
\begin{center}
\includegraphics[width=6in,angle=0,clip=true]{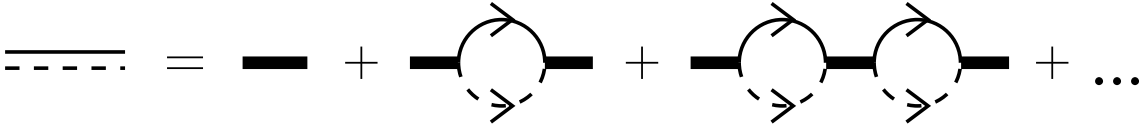}
\end{center}
\vspace*{-18pt} 
\caption{Dressing of the bare $\Lambda N$ propagator. The
dashed line indicates a $\Lambda$, the solid line a nucleon, 
the thick solid line a bare, and the 
double line a full dibaryon propagator.}
\label{fig:dress}
\end{figure}

Since the theory is nonrelativistic, all particles propagate forward in time
and the nucleon and $\Lambda$ tadpoles vanish. The propagator for
nucleons and $\Lambda$'s is simply 
\begin{equation}
\label{nucprop}
iS (p_0,\vec{p})=\frac{i}{p_0-\vec{p}^{\,2}/(2m) +i\epsilon}\,,
\end{equation}
where $m$ is the mass of the corresponding particle.
The dibaryon propagators are more complicated because of the coupling
to two-baryon states. The bare dibaryon propagators are constant, 
$i/\Delta$, but the full propagators are dressed by baryon loops
to all orders, as illustrated for the $\Lambda N$ state 
in Fig.~\ref{fig:dress}. 
%%%%%%%%%%%%%%%%%%%%%%%%%%%%%%%%%%%%%%%%%%%%%%%%%%%%%%%%%% 
%%%%%%%%%%%%%%%%%%%%%%%%%%%%%%%%%%%%%%%%%%%%%%%%%%%%%%%%%%%%%%
All diagrams in Fig.~\ref{fig:dress}
are of the same order because $2\pi\Delta/(Mg^2)=1/a\sim Q$. 
Summing the resulting geometric series leads to
\begin{equation}
\label{Uprop}
i D_j (p_0, \vec{p} ) =  \frac{2\pi}{\mu_\Lambda g_j^2}\frac{-i}
{-\gamma_j + \sqrt{-2\mu_\Lambda\left( p_0-\vec{p}^{\,2}/\left[
2(M+M_\Lambda)\right]\right) -i\epsilon}}\,,
\end{equation}
where $\mu_\Lambda=M M_\Lambda/(M+M_\Lambda)$ is the reduced 
mass of the $\Lambda N$ system, $j\in \{1,3\}$, and $\gamma_j$ is the 
corresponding pole momentum. 
Divergent loop integrals are regulated using dimensional 
regularization. The full propagator for the deuteron is \cite{BHK98,BHK00}
\begin{equation}
\label{Dprop}
i D_{d}(p_0, \vec{p} ) =  \frac{2\pi}{M g_d^2}\frac{-i}
{-\gamma_d + \sqrt{-M p_0+\vec{p}^{\,2}/4-i\epsilon}}\,,
\end{equation}
where $\gamma_d= 45.68$ MeV is the deuteron pole momentum.
The values of $\gamma_1$ and $\gamma_3$ will be discussed later, together 
with the three-body results. The scattering amplitudes are obtained 
by attaching external baryon lines to the full propagators from 
Eqs.~(\ref{Uprop},\ref{Dprop}) \cite{BHK00}.
All dependence on the bare coupling constants 
$g_d$, $g_1$, and $g_3$ cancels in observable quantities.

\section{Three-Body System}
\label{sec:3bdy}

We now apply the Lagrangian (\ref{lagd}) to the
$\Lambda NN$ system in the $J=1/2$ channel, which has the hypertriton
as a three-body bound state. 
We do not include explicit $\Lambda \leftrightarrow \Sigma$ 
conversion effects \cite{AfG89}. Such effects are accounted for
by a $\Lambda NN$ three-body force to be introduced in the following.
Explicit $\Sigma$ degrees of freedom have been integrated 
out from the effective Lagrangian. This procedure is justified
because the characteristic momentum $\gamma_3^\Lambda \ll \sqrt{M_\Lambda
(M_\Sigma -M_\Lambda)} \approx 300$ MeV \cite{Sav97}.\footnote{
See Refs.~\cite{Sav97,ORK96} for a discussion of a similar issue
in the $N\Delta$ system.}

To leading order in $QR$, only relative S-waves contribute. 
A $\Lambda NN$ state with the quantum numbers of the 
hypertriton ($J^P =\frac{1}{2}^+$) can then be constructed in three different 
ways: $(i)$ a $^3S_1$ $\Lambda N$ partial wave plus another nucleon, $(ii)$
a $^1S_0$ $\Lambda N$ partial wave plus another nucleon, and $(iii)$
a $^3S_1$ $NN$ partial wave (a deuteron) plus a $\Lambda$. 
The $^1S_0$ $NN$ partial wave does not contribute
because of isospin. The leading correction comes from the two-body
effective ranges which enter at $\order{QR}$.
Contributions from higher partial waves and the tensor force
are suppressed by at least two powers of $QR$ \cite{GBG00,HaM01b}; 
in potential models, their contribution is found to be small as well 
\cite{CJF97}.

\begin{figure}[t]
\begin{center}
\includegraphics[width=6in,angle=0,clip=true]{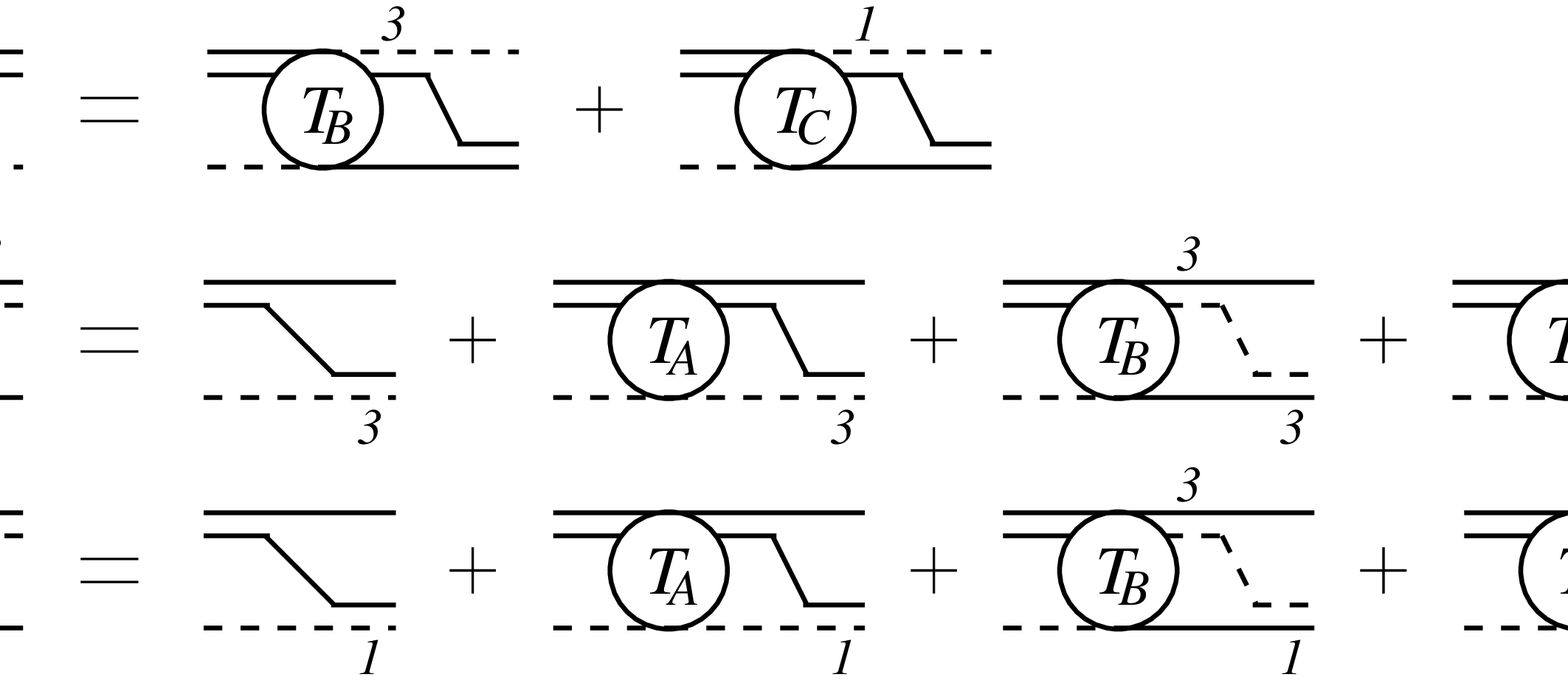}
\end{center}
\vspace*{-18pt} 
\caption{Coupled integral equations for $J=1/2$ $\Lambda d$ scattering.
Nucleons ($\Lambda$'s) are indicated by solid (dashed) lines,
respectively. The $d$ dibaryon is indicated by double solid lines,
while the $^3S_1$ ($^1S_0$) $\Lambda N$ dibaryon is indicated
by the dashed--solid double line with index 3 (1).} 
\label{fig:inteq}
\end{figure}

As a consequence, we need three three-body amplitudes, $T_A$, $T_B$, and 
$T_C$, to describe the hypertriton. 
The amplitudes satisfy the coupled integral equations shown in 
Fig.~\ref{fig:inteq}.
%%%%%%%%%%%%%%%%%%%%%%%%%%%%%%%%%%%%%%%%%%%%%%%%%%%%%%%%%% 
%%%%%%%%%%%%%%%%%%%%%%%%%%%%%%%%%%%%%%%%%%%%%%%%%%%%%%%%%%%%%%
The amplitude $T_A$ describes $J=1/2$ $\Lambda d$ scattering,
while the remaining two amplitudes have a $\Lambda d$ initial state
but go into the final states $(i)$ and $(ii)$ from above. 
Note also that there is no tree level diagram for $T_A$ because
strangeness is conserved.

The details of the derivation of the integral equations are given in 
Appendix\ \ref{app:deriv}. At a first glance, it appears that the Efimov 
physics known from the neutron-deuteron system \cite{Efi71,BHK99,BHK00}
is not present here. The masses of the $\Lambda$ and nucleon are different
and the integral equations (\ref{appx}) are not scale invariant for 
large loop momenta.
However, since the $\Lambda-N$ mass difference [characterized
by the ratio $y=(M_\Lambda-M)/(M_\Lambda+M)\approx 0.1$]
is small, the equation becomes approximately scale invariant.
Therefore, the Efimov physics will be present and
can be made manifest by expanding around the limit 
$y=0$. The error introduced by keeping only the leading order
in this expansion is at most of the order of the effective range 
corrections, which we neglect. Corrections to this limit
can be calculated within the EFT. Setting $y=0$ in Eq.~(\ref{appx}), 
we obtain:
\begin{eqnarray}
\label{aeq}
T_A (k,p) &=& -\frac{1}{\pi} \int_0^{\Lambda_c} q^2 \;dq
\left\{ L_{B}(p,q,E) T_B(k,q) -3 L_{C}(p,q,E) T_C (k,q) \right\} 
\\[10pt]
T_B (k,p) &=& -\frac{4\pi \gamma_d}{M} L_I (p,k,E)
            -\frac{1}{\pi} \int_0^{\Lambda_c} q^2 \;dq L_{A}(p,q,E) T_A(k,q)
\nonumber \\ & &
-\frac{1}{\pi} \int_0^{\Lambda_c} q^2 \;dq
\left\{ L_{B}(p,q,E) T_B(k,q) +3 L_{C}(p,q,E) T_C (k,q) \right\} 
\nonumber \\[10pt]
T_C (k,p) &=& \frac{4\pi \gamma_d}{M} L_I (p,k,E)
            +\frac{1}{\pi} \int_0^{\Lambda_c} q^2 \;dq L_{A}(p,q,E) T_A(k,q)
\nonumber \\ & &
-\frac{1}{\pi} \int_0^{\Lambda_c} q^2 \;dq
\left\{ L_{B}(p,q,E) T_B(k,q) - L_{C}(p,q,E) T_C (k,q) \right\} \,,
\nonumber
\end{eqnarray}
where $k$ ($p$) denote the incoming (outgoing) momenta in the center-of-mass
frame and $\Lambda_c$ is a momentum cutoff introduced to regulate
the integral equations. In Eqs.~(\ref{aeq}),
the outgoing momentum $p$ is taken off-shell. The total energy is 
$E = 3 k^2/(4M) - \gamma_d^2/M$.

In the limit $y=0$, the functions $L_I$ and $L_A$ are given by
\begin{eqnarray}
L_I(p,k,E)&=&\frac{1}{pk}\ln\left(
\frac{k^2+pk+p^2-ME}{k^2-pk+p^2-ME}\right)\,,\\
L_A (p,q,E)&=&\frac{1}{pq}\ln\left(
\frac{q^2+pq+p^2-ME}{q^2-pq+p^2-ME}\right)
\left[-\gamma_d+
\sqrt{\frac{3}{4}q^2-ME-i\epsilon}\right]^{-1} \,.\nonumber
\end{eqnarray}
$L_B$ ($L_C$) are identical to $L_A$ with $\gamma_d$ replaced by 
$\gamma_3$ ($\gamma_1$), respectively.
The amplitude $T_A(p,k)$ is normalized such that
\begin{equation}
T_A(k,k)=\frac{3\pi}{M}\frac{1}{k\cot\delta -ik}\,,
\end{equation}
with $\delta$ the elastic $\Lambda d$ scattering phase shift. 

The integral equations (\ref{aeq})
show the same feature as in the case of $J=1/2$ neutron-deuteron-scattering 
or three spinless bosons \cite{BHK99,BHK00}.
Without the cutoff $\Lambda_c$, their solution would not be unique
\cite{DaL63}. The origin of this nonuniqueness
can be understood by solving Eq. (\ref{aeq}) for asymptotically
large off-shell momenta $\Lambda_c \gg q\gg \gamma_d \sim k$. 
In the asymptotic limit, the inhomogeneous terms
can be neglected and the equations become independent of $k$,
\begin{eqnarray}
\widetilde{T}_A (p) &=& - \frac{2}{\sqrt{3}\pi}\int_0^\infty \frac{dq}{q}
\ln\left(\frac{p^2+pq+q^2}{p^2-pq+q^2}\right)
\left\{\widetilde{T}_B(q) -3 \widetilde{T}_C (q) \right\} 
\\
\widetilde{T}_B (p) &=& -\frac{2}{\sqrt{3}\pi} \int_0^\infty \frac{dq}{q}
\ln\left(\frac{p^2+pq+q^2}{p^2-pq+q^2}\right) \left\{ \widetilde{T}_A(q)
+ \widetilde{T}_B(q) +3 \widetilde{T}_C (q) \right\} 
\nonumber \\
\widetilde{T}_C (p) &=& +\frac{2}{\sqrt{3}\pi} \int_0^\infty \frac{dq}{q}
\ln\left(\frac{p^2+pq+q^2}{p^2-pq+q^2}\right) \left\{ \widetilde{T}_A(q)
- \widetilde{T}_B(q) + \widetilde{T}_C (q) \right\}\,,
\nonumber
\end{eqnarray}
where we have defined $\widetilde{T}_{j}(p)=p\,T_{j}(k\sim \gamma_d,p)$
for $j\in \{A,B,C\}$. Note that the asymptotic equations do not 
determine the overall normalization of the
amplitudes, which will be given from matching to the low-energy
solution of the full equation. The following discussion, however,
is independent of this normalization. We can decouple the equations by 
defining a new set of amplitudes $T_1$, $T_2$, and $T_3$ via
\begin{equation}
\left(\begin{array}{c}
{\dpst \widetilde{T}_A}\\
{\dpst \widetilde{T}_B}\\
{\dpst \widetilde{T}_C}
\end{array}\right)
=
\left(\begin{array}{ccc}
{\dpst 0\quad} & {\dpst -1-\sqrt{5}} & {\quad \dpst -1+\sqrt{5}}\\
{\dpst 3\quad} & {\dpst -1} & {\quad \dpst -1}\\
{\dpst 1\quad} & {\dpst 1} & {\quad \dpst 1}
\end{array}\right)\;
\left(\begin{array}{c}
{\dpst T_1}\\
{\dpst T_2}\\
{\dpst T_3}
\end{array}\right)\,.
\end{equation}
The new amplitudes $T_1$, $T_2$, and $T_3$ fulfill the equations
\begin{eqnarray}
\label{asympteq}
T_1 (p) &=& - \frac{4}{\sqrt{3}\pi}\int_0^\infty \frac{dq}{q}
\ln\left(\frac{p^2+pq+q^2}{p^2-pq+q^2}\right) T_1(q)
\\
T_2 (p) &=& \frac{4}{\sqrt{3}\pi} \frac{1-\sqrt{5}}{2}
\int_0^\infty \frac{dq}{q}
\ln\left(\frac{p^2+pq+q^2}{p^2-pq+q^2}\right) T_2 (q) 
\nonumber \\
T_3 (p) &=& \frac{4}{\sqrt{3}\pi} \frac{1+\sqrt{5}}{2}
\int_0^\infty \frac{dq}{q}
\ln\left(\frac{p^2+pq+q^2}{p^2-pq+q^2}\right) T_3 (q) \,.
\nonumber
\end{eqnarray}
All three equations are scale invariant and symmetric under
the inversion $q \to 1/q$. In Ref.~\cite{DaL63}, 
it was shown that an equation of the type
\begin{equation}
f(p) = \frac{4\lambda}{\sqrt{3}\pi}
\int_0^\infty \frac{dq}{q}
\ln\left(\frac{p^2+pq+q^2}{p^2-pq+q^2}\right) f(q)
\end{equation}
has a unique solution in the form of a power law if $\lambda <
\lambda_c=3\sqrt{3}/(4\pi)\approx 0.4135$ (see also 
Refs.~\cite{BHK99,BHK00}). This condition is clearly
satisfied for the first two equations. The equation for $T_3$,
however, has $\lambda=(1+\sqrt{5})/2\approx 1.62$.
As a consequence, there are two linearly independent complex solutions 
$T_3(k,p)=p^{\pm i s_1}$, where $s_1=1.35322$. 
The relative phase between these two solutions is not
determined by Eq.~(\ref{asympteq}). For a finite cutoff $\Lambda_c$ this
phase is fixed but strongly depends on $\Lambda_c$. This $\Lambda_c$
dependence can absorbed by adding a one-parameter
three-body force in the equation for $T_3$,
\begin{equation}
\label{eq:asy3bdy}
T_3 (p) = \frac{4}{\sqrt{3}\pi} \frac{1+\sqrt{5}}{2}
\int_0^{\Lambda_c} \frac{dq}{q}\left\{
\ln\left(\frac{p^2+pq+q^2}{p^2-pq+q^2}\right) 
+2 H(\Lambda_c) \frac{pq}{\Lambda_c^2} \right\} T_3 (q) \,,
\nonumber
\end{equation}
where $H(\Lambda_c)$ is a dimensionless function of the cutoff.
In Refs.~\cite{BHK99,BHK00}, it was shown that if
the three-body force runs with cutoff as
\begin{equation}
\label{runH}
H(\Lambda_c)=-    \frac{\sin(s_1\ln({\Lambda_c}/{\Lambda_*})-
                   {\rm arctg}(1/s_1))}
                 {\sin(s_1 \ln({\Lambda_c}/{\Lambda_*})+
                   {\rm arctg}(1/s_1))}\,,
\end{equation}
all low-energy observables are independent of $\Lambda_c$.
The renormalization group evolution of $H(\Lambda_c)$ is governed
by a limit cycle. In particular, if the cutoff is increased by a factor 
of $\exp(\pi/s_1)\approx 10.2$, $H(\Lambda_c)$ returns to its
original value. $\Lambda_*$ is a dimensionful 
parameter that determines the asymptotic phase of the off-shell 
amplitude \cite{BHK99,BHK00}. It has to be fixed from a three-body
observable. Once its value is known, all other observables can
be predicted.\footnote{One can either specify the dimensionless
coupling $H$ at a specific cutoff $\Lambda_c$ or the dimensionful
low-energy parameter $\Lambda_*$. This is similar to dimensional
transmutation in QCD.}
Note, however, that $\Lambda_*$ is only determined up to factors of
$\exp(\pi/s_1)$.

Formally, the three-body force term is obtained by adding
a nonderivative three-body contact term to the effective Lagrangian
(\ref{lagd}). The form of this term can be determined by starting from 
a general structure and matching the unknown coefficients 
to reproduce Eq.~(\ref{eq:asy3bdy}). The exact expression for this 
three-body term is given in Appendix\ \ref{app:3bdy}. It is 
interesting to note that the value $s_1\approx 1.35$ differs from 
$s_0 \approx 1.006$ which characterizes
the three-nucleon force in the triton channel \cite{BHK00}. 
This does not lead to a contradiction because the hypertriton channel
($I=0$) and the triton channel ($I=1/2$) have different isospin.

For the numerical studies, it is useful to write a renormalized 
integral equation by choosing a special cutoff,
\begin{eqnarray}
\label{zero}
\Lambda_n &=& \Lambda_* \exp\left[\frac{1}{s_1}\left(n \pi +
\arctan\left(\frac{1}{s_1}\right) \right) \right]\,,
\end{eqnarray} 
with $n>0$ an integer, at which the the three-body term proportional to
$H$ in Eq.~(\ref{eq:asy3bdy}) vanishes \cite{HaM01}. 
This procedure does not remove
the dependence on the three-body parameter $\Lambda_*$ but rather
moves its dependence into the cutoff. All observables are independent
of $n$, but it should be chosen such that $\Lambda_n \gg \gamma_d$
and finite cutoff corrections of order $\gamma_d/\Lambda_n$ can
be neglected. In the following, we will be using Eqs.~(\ref{aeq}) with
$\Lambda_c =\Lambda_1$. The $\Lambda d$ scattering amplitude is 
obtained by numerically solving Eqs.~(\ref{aeq}) for the 
desired value of the total energy. The hypertriton binding 
energy can be extracted by solving Eqs.~(\ref{aeq}) for negative 
energies and locating the position of the bound state pole
in $T_A$.

\section{Results}
\label{sec:res}

In order to apply the EFT to the $\Lambda NN$ system, we have to fix the 
parameters in Eqs.~(\ref{aeq}).
The deuteron binding momentum $\gamma_d = 45.68$ MeV is known
very well. As mentioned above, the low-energy parameters
for the $\Lambda N$ S-waves cannot be extracted unambiguously
from a model-independent analysis of low-energy scattering data 
\cite{Ale68,Sec68}. It is known, however, that the $\Lambda N$
system is unbound, which requires $\gamma_1$ and $\gamma_3$ to
be negative. The errors in the low-energy parameters extracted from 
extrapolations using hyperon-nucleon interaction potentials 
\cite{NRS73,RSY99,HHS89,RHS94} are not well known and can only be 
estimated from the variation with different potentials.
Therefore, we do not take the extracted parameters as input. Rather,
we vary the $\Lambda N$ pole momenta over the range of validity
of the EFT  and study the implications on the hypertriton and
$\Lambda d$ scattering.

In principle, we have three unknown parameters: $\gamma_1$, $\gamma_3$, 
and the three-body parameter $\Lambda_*$. However, we will first address 
the simpler scenario $\gamma_1=\gamma_3$, which follows from the $SU(6)$ 
spin-flavor symmetry of the quark model \cite{AkD65,BaR65}. 
Such an $SU(6)$ symmetry also emerges in the large-$N_c$ limit of 
QCD \cite{KaS96}. Symmetry violating corrections are suppressed
by $1/N_c$ and the strange quark mass. Since both isospin and strangeness
of the $^3S_1$ and $^1S_0$ $\Lambda N$ partial waves are the same, 
the difference between $\gamma_1$ and $\gamma_3$ is a pure
spin splitting effect. In most hyperon-nucleon potentials this effect
is of the order 30\% \cite{NRS73,RSY99,HHS89,RHS94}. 
Neglecting this splitting leaves us with two parameters:
$\gamma_i(=\gamma_1=\gamma_3)$ and $\Lambda_*$. 
In the following, we vary these two parameters and study 
the constraints that follow from a correct description of the hypertriton. 
We will also elucidate the consequences for $\Lambda d$ scattering.

\begin{figure}[t]
\begin{center}
\includegraphics[width=4.5in,angle=0,clip=true]{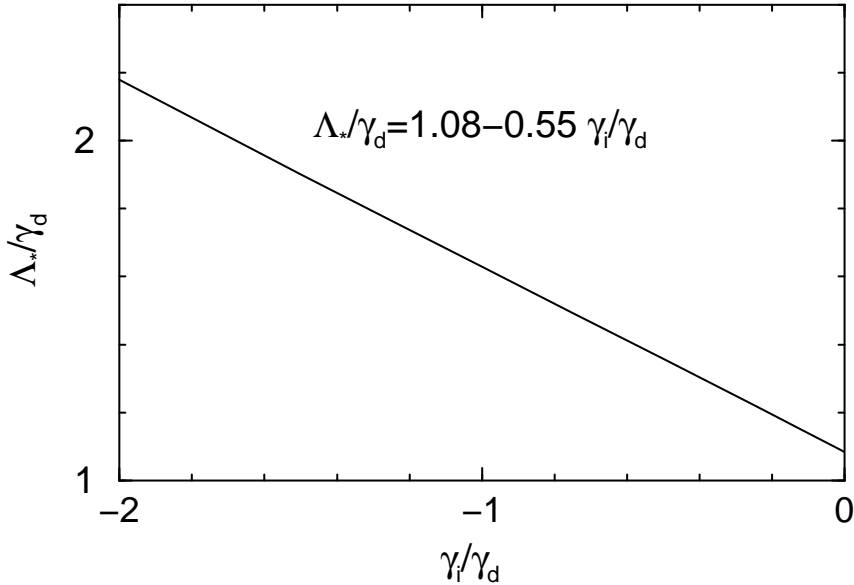}
\end{center}
\vspace*{-18pt} 
\caption{Values of  $\Lambda_*/\gamma_d$ and  $\gamma_i/\gamma_d$ that
lead to the correct binding energy for the hypertriton.}
\label{fig:gamils}
\end{figure}

First, we study the constraints on the parameters $\gamma_i$ and 
$\Lambda_*$ by the requirement that the hypertriton binding energy be
reproduced. An interesting question is whether there are certain values of 
$\gamma_i$ and $\Lambda_*$ that are incompatible with the hypertriton
data. Furthermore, it is possible that the approximation $\gamma_1=\gamma_3$ 
proves too restrictive. In Fig.~\ref{fig:gamils},
we plot $\Lambda_*$ as a function of $\gamma_i$ with the
%%%%%%%%%%%%%%%%%%%%%%%%%%%%%%%%%%%%%%%%%%%%%%%%%%%%%%%%%% 
%%%%%%%%%%%%%%%%%%%%%%%%%%%%%%%%%%%%%%%%%%%%%%%%%%%%%%%%%%%%%%
requirement that the hypertriton binding energy is reproduced.
We vary $\gamma_i$ from $0$ to $-2\gamma_d$, which is about where
corrections of $\order{QR}$ are expected to become important.
Positive values of $\gamma_i$ are excluded because the $\Lambda N$
system is not bound.
All allowed values fall on a curve with a linear dependence on
$\gamma_i$,
\begin{equation}
\Lambda_*/\gamma_d =1.08-0.55 \gamma_i/\gamma_d\,.
\end{equation}
For every value of $\gamma_i$ in the interval a corresponding value of
$\Lambda_*$ that reproduces the hypertriton can be found.
Clearly, the exact value of $\gamma_i$ and therefore
the $\Lambda N$ scattering lengths cannot be 
determined from the hypertriton binding energy. 
This is in contrast to Ref.~\cite{CJF97}, where
it was found that with standard hyperon-nucleon potentials,
the correct binding energy for the hypertriton 
cannot be reproduced if the singlet $\Lambda N$ scattering
length differs more than 10\% from 1.85 fm.

\begin{figure}[t]
\begin{center}
\includegraphics[width=4.5in,angle=0,clip=true]{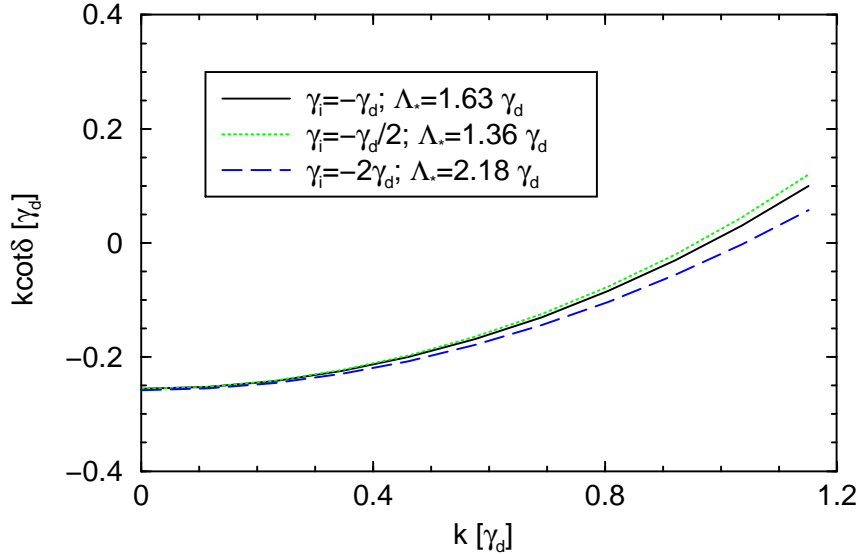}
\end{center}
\vspace*{-18pt} 
\caption{$\Lambda d$ scattering phase shifts for different values of 
$\gamma_i$ and $\Lambda_*$ that lead to the correct hypertriton
binding energy (all quantities in units of $\gamma_d$).}
\label{fig:htrib}
\end{figure}

In Fig.~\ref{fig:htrib}, we compare the scattering phase
shifts for $\Lambda d$ scattering in the hypertriton channel
for different points on the curve in Fig.~\ref{fig:gamils}. 
Fig.~\ref{fig:htrib} shows three curves with
$\gamma_i$ chosen to be equal in magnitude to $\gamma_d$ as
well as a factor of two smaller and larger.
The phase shifts agree very well for small momenta around threshold
and begin to differ slightly close to the deuteron breakup
threshold. This implies that low-energy $\Lambda d$ scattering
does not give additional constraints for the values of $\gamma_i$ 
and $\Lambda_*$. Consequently, neither the hypertriton binding energy
nor the low-energy $\Lambda d$ phase shifts can be used to fix the 
exact value of $\gamma_i$. Such information can only come from 
low-energy $\Lambda N$ scattering. 
%%%%%%%%%%%%%%%%%%%%%%%%%%%%%%%%%%%%%%%%%%%%%%%%%%%%%%%%%% 
%%%%%%%%%%%%%%%%%%%%%%%%%%%%%%%%%%%%%%%%%%%%%%%%%%%%%%%%%%%%%%
On the other hand, this observation
allows for a prediction of the low-energy $\Lambda d$ phase shifts
[given by the curves in Fig.~\ref{fig:htrib}] independent
of the exact values for $\gamma_i$ and $\Lambda_*$.
This is a direct consequence of the shallowness of the hypertriton.
Extracting the low-energy doublet $\Lambda d$ scattering parameters
from the phase shifts, we find for the scattering 
length $a_{\Lambda d}$ and effective range $r_{\Lambda d}$:
\begin{equation}
\label{eq:ladeff}
a_{\Lambda d}=(16.8^{+4.4}_{-2.4}) \mbox{ fm} \qquad \mbox{and} \qquad
r_{\Lambda d}=(2.3 \pm 0.3) \mbox{ fm}\,,
\end{equation}
where the errors are due to the uncertainty in the hypertriton
binding energy. Our value for the scattering length is in good agreement 
with Ref.~\cite{CJF97}, while the effective range is about 1 fm smaller.

\begin{figure}[t]
\begin{center}
\includegraphics[width=4.5in,angle=0,clip=true]{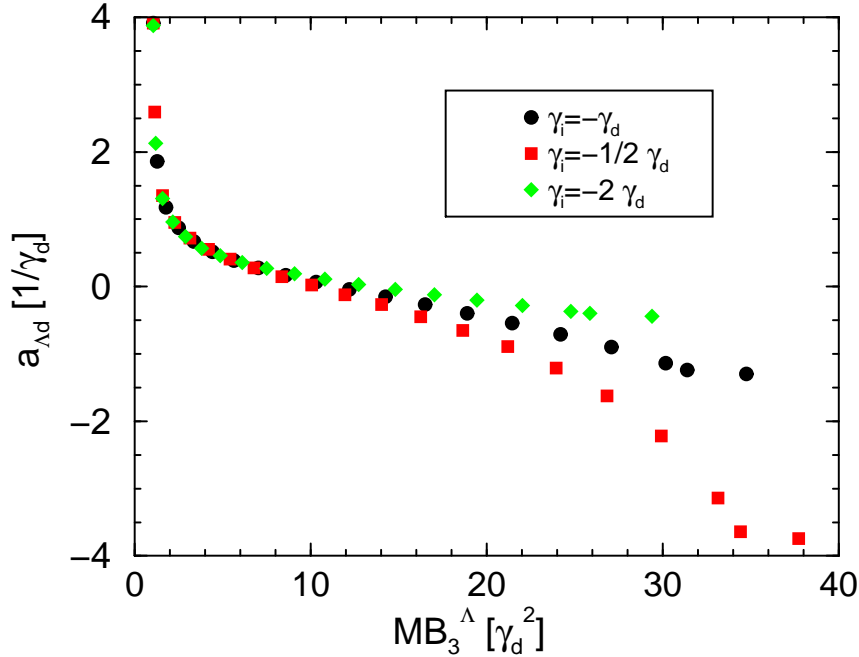}
\end{center}
\vspace*{-18pt} 
\caption{Phillips line for different values of $\gamma_i$ (all quantities 
in units of $\gamma_d$).}
\label{fig:phil}
\end{figure}

It is also interesting to look at the \lq\lq Phillips line'',
which describes the correlation between the $\Lambda d$ scattering
length $a_{\Lambda d}$ and the hypertriton binding energy $B_3^\Lambda$. 
The Phillips line was observed in the neutron-deuteron system, where
the results of calculations of the doublet $nd$ scattering 
length and the triton binding energy using different $NN$ potentials
fall on a line when plotted in a plane \cite{Phi68}. A similar line 
exists for the hypertriton (cf. Fig.~7 in Ref.~\cite{FeJ01}).
In the EFT framework, the Phillips line is parametrized by variations 
in the three-body  parameter $\Lambda_*$ \cite{BHK99,BHK00}. 
In Fig.~\ref{fig:phil}, we show the
%%%%%%%%%%%%%%%%%%%%%%%%%%%%%%%%%%%%%%%%%%%%%%%%%%%%%%%%%% 
%%%%%%%%%%%%%%%%%%%%%%%%%%%%%%%%%%%%%%%%%%%%%%%%%%%%%%%%%%%%%%
Phillips line for the three values of $\gamma_i$ from Fig.~\ref{fig:htrib}. 
For small $B_3^\Lambda$, the different Phillips lines
coincide exactly (the physical hypertriton corresponds
to $MB_3^\Lambda \approx 1.06\,\gamma_d^2$) and deviate from each other 
only at extremely large binding energies. However, the EFT is 
expected to break down when the three-body binding momentum 
$\gamma_3^\Lambda$ from Eq.~(\ref{eq:3bdymom}) becomes of the order of 
the pion mass. This is the case for $MB_3^\Lambda \simge 8 \gamma_d^2$.
Consequently, the region where the Phillips lines differ is
unphysical. For all practical purposes the Phillips line is independent
of $\gamma_i$. 

The characteristic momentum that sets the scale of corrections to 
the large-scattering-length approximation is given by  the binding momentum 
of the hypertriton. Using the measured value for $B_3^\Lambda$
and Eq.~(\ref{eq:3bdymom}), we obtain $\gamma_3^\Lambda \approx 
0.3\,\gamma_d$. This momentum can be used to estimate the contribution
of the effective range term in the two-body amplitudes which is the
leading correction at $\order{QR}$.
For our approximation to be accurate, we need 
$|\gamma| \gg |r (\gamma_3^\Lambda)^2/2|$ in each of the two-body partial 
waves. Since $(\gamma_3^\Lambda)^2/2 \approx 0.05\,\gamma_d^2$ is extremely
small, the effective range term gives only a small correction even when 
the scattering length and effective range are of the same order. 

We have also studied the more general case $\gamma_1 \neq \gamma_3$.
The conclusions in this case are unchanged: requiring the correct 
binding energy of the hypertriton does not constrain the specific values
of $\gamma_1$ and $\gamma_3$ as long as they are within the range of 
validity of the EFT ($0 > \gamma_1,\gamma_3 \simge -2\gamma_d$). 
The results for $\Lambda d$ scattering confirm Fig.~\ref{fig:htrib}
and Eq.~(\ref{eq:ladeff}).

The insensitivity to the precise values of $\gamma_1$ and $\gamma_3$ and
the small separation energy of the hypertriton into a deuteron and a 
$\Lambda$ ($B^\Lambda =0.13$ MeV) can be exploited even further: it is
possible to write an EFT with only deuterons and $\Lambda$'s as degrees
of freedom \cite{Bira01}. Effects from the deuteron structure 
and $\Lambda N$ interaction can be included perturbatively via local
operators.

To summarize, we have discussed the hypertriton and doublet $\Lambda d$
scattering in an EFT for large two-body scattering lengths.
Due to the small binding energy of the hypertriton, our analysis 
is expected to hold even if the $\Lambda N$ low-energy parameters
are $\order{1/m_\pi}$ as extracted from potential models.
As in the triton case, consistent renormalization 
requires a one-parameter three-body force at leading order whose 
renormalization group evolution is governed by a limit cycle. 
The period of the limit cycle
$\exp(\pi/s_1)\approx 10.2$ is a factor two smaller than for the triton.
In contrast to standard potential models where the binding
energy of the hypertriton is very sensitive to the singlet $\Lambda N$
scattering length \cite{CJF97}, the three-body force can always
compensate for changes in the low-energy $\Lambda N$ parameters.
While the $\Lambda N$ low-energy 
parameters cannot be fixed by requiring the correct binding energy 
for the hypertriton, the shallowness of the hypertriton still allows for
a unique prediction of low-energy doublet $\Lambda d$ scattering.
Using the experimental binding energy of the hypertriton as input,
we find for the scattering length $a_{\Lambda d}=(16.8^{+4.4}_{-2.4})$ fm and 
for the  effective range $r_{\Lambda d}=(2.3\pm 0.3)$ fm.
In the future, it would be interesting to go to higher orders and
apply EFT methods to a wider class of observables and other halo nuclei. 
Work in this direction is in progress \cite{BHK01}.
\section*{Acknowledgements}
I would like to thank E.\ Braaten, R.J.\ Furnstahl, U.\ van Kolck, 
T.\ Mehen, and A.\ Parre{\~n}o for useful discussions. 
This research was supported by the U.S. National Science 
Foundation under Grant Nos.\ PHY-9800964 and PHY-0098645.
\begin{appendix}
\section{Derivation of three-body equation}
\label{app:deriv}
In this Appendix, we give some details of the derivation of the 
integral equation (\ref{aeq}) for the general case $M_\Lambda \neq
M$.
From the Lagrangian (\ref{lagd}) and the Feynman diagrams in
Fig.~\ref{fig:inteq}, we obtain\footnote{See Refs.~\cite{BHK00,GBG00}
for details on the evaluation of the Feynman diagrams.}
\begin{eqnarray}
\label{app:1}
t_A^{ij}(\vec{k},\vec{p})_{\alpha\beta} &=& g_d g_3 \int
 \frac{d^3 q}{(2\pi)^3}\; t_B^{ik}(\vec{k},\vec{q})_{\alpha\gamma}\,
 \frac{(\sigma_j \sigma_k)_{\gamma\beta}\; D_3 (E-
 \frac{q^2}{2M},\vec{q}) }{E-\frac{p^2}{2M_\Lambda}-\frac{q^2}{2M}
 -\frac{(\vec{q}+\vec{p})^2}{2 M}+i\epsilon}
\\
& & + g_d g_1 \int
 \frac{d^3 q}{(2\pi)^3}\; t_C^{i}(\vec{k},\vec{q})_{\alpha\gamma}\,
 \frac{(\sigma_j)_{\gamma\beta}\; D_1 (E-
 \frac{q^2}{2M},\vec{q}) }{E-\frac{p^2}{2M_\Lambda}-\frac{q^2}{2M}
 -\frac{(\vec{q}+\vec{p})^2}{2 M}+i\epsilon}
\nonumber \\
t_B^{ij}(\vec{k},\vec{p})_{\alpha\beta} &=& 
-2g_d g_3 \frac{(\sigma_j \sigma_i)_{\alpha\beta}}
  {E-\frac{k^2}{2M_\Lambda}-\frac{p^2}{2M}-\frac{(\vec{k}+\vec{p})^2}
  {2 M} +i\epsilon}
\\
& &+2 g_d g_3 \int
 \frac{d^3 q}{(2\pi)^3}\; t_A^{ik}(\vec{k},\vec{q})_{\alpha\gamma}\,
 \frac{(\sigma_j \sigma_k)_{\gamma\beta}\; D_d (E-
 \frac{q^2}{2M_\Lambda},\vec{q}) }{E-\frac{p^2}{2M}-\frac{q^2}{2M_\Lambda}
 -\frac{(\vec{q}+\vec{p})^2}{2 M}+i\epsilon}
\nonumber\\
&& +g_3^2 \int
 \frac{d^3 q}{(2\pi)^3}\; t_B^{ik}(\vec{k},\vec{q})_{\alpha\gamma}\,
 \frac{(\sigma_j \sigma_k)_{\gamma\beta}\; D_3 (E-
 \frac{q^2}{2M},\vec{q}) }{E-\frac{p^2+q^2}{2M}
 -\frac{(\vec{q}+\vec{p})^2}{2 M_\Lambda}+i\epsilon}
\nonumber \\
& & -g_1 g_3 \int
 \frac{d^3 q}{(2\pi)^3}\; t_C^{i}(\vec{k},\vec{q})_{\alpha\gamma}\,
 \frac{(\sigma_j)_{\gamma\beta}\; D_1 (E-
 \frac{q^2}{2M},\vec{q}) }{E-\frac{p^2+q^2}{2M}
 -\frac{(\vec{q}+\vec{p})^2}{2 M_\Lambda}+i\epsilon}
\nonumber\\
t_C^{i}(\vec{k},\vec{p})_{\alpha\beta} &=& 
-2g_d g_1 \frac{(\sigma_i)_{\alpha\beta}}
  {E-\frac{k^2}{2M_\Lambda}-\frac{p^2}{2M}-\frac{(\vec{k}+\vec{p})^2}
  {2 M} +i\epsilon}
\\
& &+2 g_d g_1 \int
 \frac{d^3 q}{(2\pi)^3}\; t_A^{ik}(\vec{k},\vec{q})_{\alpha\gamma}\,
 \frac{(\sigma_k)_{\gamma\beta}\; D_d (E-
 \frac{q^2}{2M_\Lambda},\vec{q}) }{E-\frac{p^2}{2M}-\frac{q^2}{2M_\Lambda}
 -\frac{(\vec{q}+\vec{p})^2}{2 M}+i\epsilon}
\nonumber\\
&& -g_1 g_3 \int
 \frac{d^3 q}{(2\pi)^3}\; t_B^{ik}(\vec{k},\vec{q})_{\alpha\gamma}\,
 \frac{(\sigma_k)_{\gamma\beta}\; D_3 (E-
 \frac{q^2}{2M},\vec{q}) }{E-\frac{p^2+q^2}{2M}
 -\frac{(\vec{q}+\vec{p})^2}{2 M_\Lambda}+i\epsilon}
\nonumber \\
& & +g_1^2 \int
 \frac{d^3 q}{(2\pi)^3}\; t_C^{i}(\vec{k},\vec{q})_{\alpha\gamma}\,
 \frac{\delta_{\gamma\beta}\; D_1 (E-
 \frac{q^2}{2M},\vec{q}) }{E-\frac{p^2+q^2}{2M}
 -\frac{(\vec{q}+\vec{p})^2}{2 M_\Lambda}+i\epsilon}\,,
\nonumber
\end{eqnarray}
where $i,j,k$ are vector indices and $\alpha,\beta,\gamma$ are 
spinor indices in spin space. $k (p)$ are the incoming (outgoing)
momenta in the center-of-mass  frame and  $E = k^2/(4M) - \gamma_d^2/M + 
k^2/(2M_\Lambda)$ is the total energy. The momentum $p$ is taken 
off-shell. The isospin dependence of the amplitudes 
$t_B$ and $t_C$ has been absorbed into the amplitudes via
the definition
\begin{eqnarray}
t_{B}^{ij}(\vec{k},\vec{p})_{\alpha\beta} &=&
t_{B}^{ij}(\vec{k},\vec{p})_{\alpha\beta}^{eb} (\tau_2)_{be}\,, 
\end{eqnarray}
and similarly for $t_C$, where $e,b$ are isospinor indices.
Next we project onto total angular momentum $J=1/2$ by using 
the definitions
\begin{eqnarray}
t_{A/B}(\vec{k},\vec{p})\, \delta_{\alpha\beta} &=&
\frac{1}{3} (\sigma_i)_{\alpha\alpha'}
t_{A/B}^{ij}(\vec{k},\vec{p})_{\alpha'\beta'} (\sigma_j)_{\beta'\beta}
\\
t_{C}(\vec{k},\vec{p})\, \delta_{\alpha\beta} &=&
\frac{1}{3} (\sigma_i)_{\alpha\alpha'}
t_{C}^{i}(\vec{k},\vec{p})_{\alpha'\beta}\,,
\nonumber
\end{eqnarray}
and projecting on relative S-waves. We also have to account
for the wave function renormalization of the deuteron,
\begin{equation}
Z_d^{-1}=i\frac{\partial}{\partial p_0}\left.(iD_d(p_0,\vec{p}))^{-1}
\right|_{p_0=-\gamma_d^2/M,\, \vec{p}=0}=
\frac{M^2 g_d^2}{4\pi\gamma_d}\,.
\end{equation}
Defining 
\begin{eqnarray}
T_A (k,p) &=& Z_d t_A(k,p)\\
T_B (k,p) &=& Z_d \frac{g_d}{g_3} t_B(k,p) \nonumber\\
T_C (k,p) &=& Z_d \frac{g_d}{g_1} t_C(k,p)\,, \nonumber
\end{eqnarray}
we finally obtain the integral equations
\begin{eqnarray}
\label{appx}
T_A (k,p) &=& -\frac{1}{\pi(1-y)}\int_0^{\Lambda_c} q^2 \;dq
\left\{ \widetilde{L}_{B}(p,q,E) T_B(k,q) 
        -3 \widetilde{L}_{C}(p,q,E) T_C (k,q) \right\} 
\\
T_B (k,p) &=& -\frac{4\pi \gamma_d}{M} L_I (p,k,E)
          -\frac{1}{\pi} \int_0^{\Lambda_c} q^2 \;dq L_{A}(p,q,E) T_A(k,q)
\nonumber\\ & &
- \frac{1}{\pi(1-y)}\int_0^{\Lambda_c} q^2 \;dq
\left\{ L_{B}(p,q,E) T_B(k,q) +3 L_{C}(p,q,E) T_C (k,q) \right\} 
\nonumber \\
T_C (k,p) &=& \frac{4\pi \gamma_d}{M} L_I (p,k,E)
          +\frac{1}{\pi} \int_0^{\Lambda_c} q^2 \;dq L_{A}(p,q,E) T_A(k,q)
\nonumber\\ & &
- \frac{1}{\pi(1-y)} \int_0^{\Lambda_c} q^2 \;dq
\left\{ L_{B}(p,q,E) T_B(k,q) - L_{C}(p,q,E) T_C (k,q) \right\} 
\nonumber
\end{eqnarray}
where $\Lambda_c$ is a momentum cutoff and
$y=(M_\Lambda-M)/(M_\Lambda+M)\approx 0.1$ characterizes the
$\Lambda-N$ mass difference.
The functions $L_I$, $L_A$, $L_B$, and $\widetilde{L}_B$ are given by
\begin{eqnarray}
L_I(p,k,E)&=&\frac{1}{pk}\ln\left(
\frac{k^2/(1+y)+pk+p^2-ME}{k^2/(1+y)-pk+p^2-ME}\right)\,,\\
L_A(p,q,E)&=&\frac{1}{pq}\ln\left(
\frac{q^2/(1+y)+pq+p^2-ME}{q^2/(1+y)-pq+p^2-ME }\right) \nonumber\\
& &\times\left[-\gamma_d+
\sqrt{ q^2 \,(3-y)/(1+y)/4 -ME -i\epsilon}\right]^{-1} \,,\nonumber \\
L_B(p,q,E)&=&\frac{1}{pq}\ln\left(
\frac{q^2+p^2+pq(1-y)-M E(1+y)}{q^2+p^2-pq(1-y)-M E(1+y)}\right)
\nonumber \\
& &\times\left[-\gamma_3+
\sqrt{ q^2 \left(3+2 y -y^2 \right)/4
  -ME(1+y) -i\epsilon}\right]^{-1} \,,\nonumber\\
\widetilde{L}_B(p,q,E)&=&\frac{1}{pq}\ln\left(
\frac{q^2+pq+p^2/(1+y)-M E}{q^2-pq+p^2/(1+y)-M E}\right)
\nonumber \\
& &\times\left[-\gamma_3+
\sqrt{ q^2 \left(3+2 y -y^2 \right)/4
  -ME(1+y) -i\epsilon}\right]^{-1} \,.\nonumber
\end{eqnarray}
$L_C$ $(\widetilde{L}_C)$ is identical to $L_B$ $(\widetilde{L}_B)$
with $\gamma_3$ replaced by $\gamma_1$.
\section{Three-body force term}
\label{app:3bdy}
The three-body term proportional to $H(\Lambda_c)$ in Eq.~(\ref{eq:asy3bdy})
can be obtained by adding a local, nonderivative three-body contact
term to the Lagrangian (\ref{lagd}). This term is most easily determined 
by starting from a general structure and matching the coefficients
of the individual terms to reproduce the three-body term in 
Eq.~(\ref{eq:asy3bdy}). This leads to,
\begin{eqnarray}
\label{eq:3bod}
{\cal L}_3 &=&  \frac{MH(\Lambda_c )}{20\Lambda_c^2}\bigg\{ 
 16\sqrt{5}\, g_d^2 \left[ d_l^* \Lambda_\alpha (\sigma_l 
 \sigma_{l'})_{\alpha\beta} \Lambda^*_\beta d_{l'} \right] 
\\
&-& (5+3\sqrt{5})\left( g_3^2 \left[ (u^3_l)^*_a N_{\alpha a} 
    (\sigma_{l} \sigma_{l'})_{\alpha\beta}
    N^*_{\beta b} (u^3_{l'})_b \right] 
 -  3 g_1^2 \left[ (u^1)^*_a N_{\alpha a} N^*_{\beta b}(u^1)_b \right]\right)
\nonumber \\  
&+& 2(5+\sqrt{5})g_d g_3 \left[ d_l^* N_{\alpha a} 
    (\sigma_{l} \sigma_{l'})_{\alpha\beta}
    \Lambda^*_{\beta} (\tau_2)_{ab} (u^3_{l'})_b +\mbox{H.c.}
    \right] 
\nonumber \\
&-& (5+3\sqrt{5})g_1 g_3 \left[ (u^3_l)^*_a N_{\alpha a} 
    (\sigma_{l})_{\alpha\beta}
    N^*_{\beta b} (u^1)_b +\mbox{H.c.} \right] 
\nonumber\\
&+& 2(5+\sqrt{5})g_d g_1 \left[ d_l^* N_{\alpha a} 
    (\sigma_{l})_{\alpha\beta}
    \Lambda^*_{\beta} (\tau_2)_{ab} (u^1)_b +\mbox{H.c.}
    \right] \bigg\}\,,
\nonumber
\end{eqnarray}
where $\alpha,\beta$ ($a,b$) are spinor (isospinor) indices and
$l,l'$ are vector indices. 
Eq.~(\ref{eq:3bod}) represents a contact three-body force written in
terms of dibaryon, nucleon, and $\Lambda$ fields but to leading order
is equivalent to a three-baryon contact force. This becomes obvious if
the dibaryon fields are integrated out by performing a Gaussian path 
integral \cite{BHK99,BHK00,GBG00}.
\end{appendix}

\end{document}